%% file: ICRC2023_main_text_IceCube.tex
\documentclass[a4paper,11pt]{article}

\usepackage{pos}
\usepackage{lipsum} %package to generate placeholder text in the following

\title{Enhanced Starting Track Real-time Stream for IceCube}

\ShortTitle{Enhanced Starting Track Real-time Stream for IceCube}

\author{The IceCube Collaboration \\{\normalsize \normalfont(a complete list of authors can be found at the end of the proceedings)}\\}

\emailAdd{jsosborn2@wisc.edu}

\abstract{

IceCube real-time alerts allow for rapid follow-up observations of likely astrophysical neutrino events, enabling searches for multi-messenger counterparts. The Enhanced Starting Track Real-time Stream (ESTReS) is a real-time extension of the Enhanced Starting Track Event Selection (ESTES), a high astrophysical purity muon-neutrino sample recently used by IceCube to measure the astrophysical diffuse flux. A set of computationally cheap cuts allows us to run a fast filter in seconds. This online filter selects about 100 events per day to be sent to Madison, WI via satellite where the full ESTES event selection is applied within minutes. Events that pass the final set of cuts (ESTReS + ESTES) will be sent out as real-time alerts to the broader astrophysical community. ESTReS’s unique contribution to the current real-time alerts will be events in the southern sky in the 5 TeV - 100 TeV range. We expect about 10.3 events per year which average 50\% astrophysical purity. In this talk I will report the status of the ESTReS alert stream in the context of the IceCube real-time program.
\vspace{4mm}
{\bfseries Corresponding authors:} Jesse Osborn$^{1*}$, Sarah Mancina$^{1,2}$, Manuel Silva$^{1}$\\
{$^{1}$ \itshape University of Wisconsin, Madison}\\
{$^{2}$ \itshape Universit{\`a} Degli Studi di Padova}\\
$^*$ Presenter

\ConferenceLogo{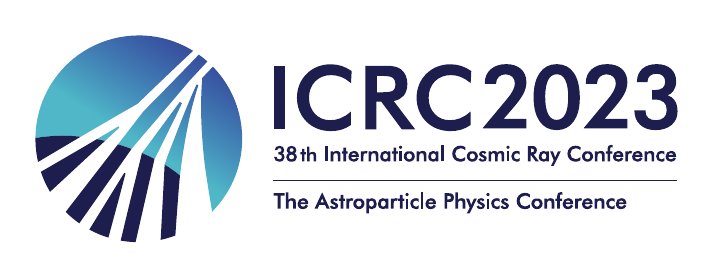}

\FullConference{The 38th International Cosmic Ray Conference (ICRC2023)\\ 26 July -- 3 August, 2023\\ Nagoya, Japan}
}

\begin{document}

\maketitle

\section{Introduction}\label{sec1}

IceCube is a cubic-kilometer neutrino detector located at the South Pole \cite{IceCubeDetector} which completed construction in December 2010. It consists of digital-optical-modules (DOMs) containing photomultiplier tubes buried between 1450 m and 2450 m into the ice. It optically observes the Cherenkov radiation that is emitted by charged particles which are produced in the weak interactions between incoming neutrinos and the surrounding ice.

Since 2016 (and improved in 2019), IceCube has been sending out alerts for possible astrophysical neutrinos that are detected in real-time to the multi-messenger astrophysics community within minutes of detection. This real-time alert stream is made up of several different event selections which are detailed in Ref. \cite{IceCubeRealtime}. We present a new event selection which will predominantly provide events from the southern sky and with reconstructed energies of 5-100 TeV. These events are currently not well covered by IceCube's real-time stream. We will be able to double the number of IceCube Gold (50\% average astrophysical purity) alerts sent out per year, from $\sim$10 to $\sim$20.

In Sec 2, we describe the new event selection for alerts, and the evaluated event rates using Monte Carlo (MC) and 10.3 years of archival data. In Sec 3, we describe the calculation of the signalness variable per event. In Sec 4, we detail the information that will be sent out as alerts to the community.

\section{Event Selection}\label{sec2}

\begin{figure}[h!]
    \centering
    \includegraphics[width=0.49\textwidth]{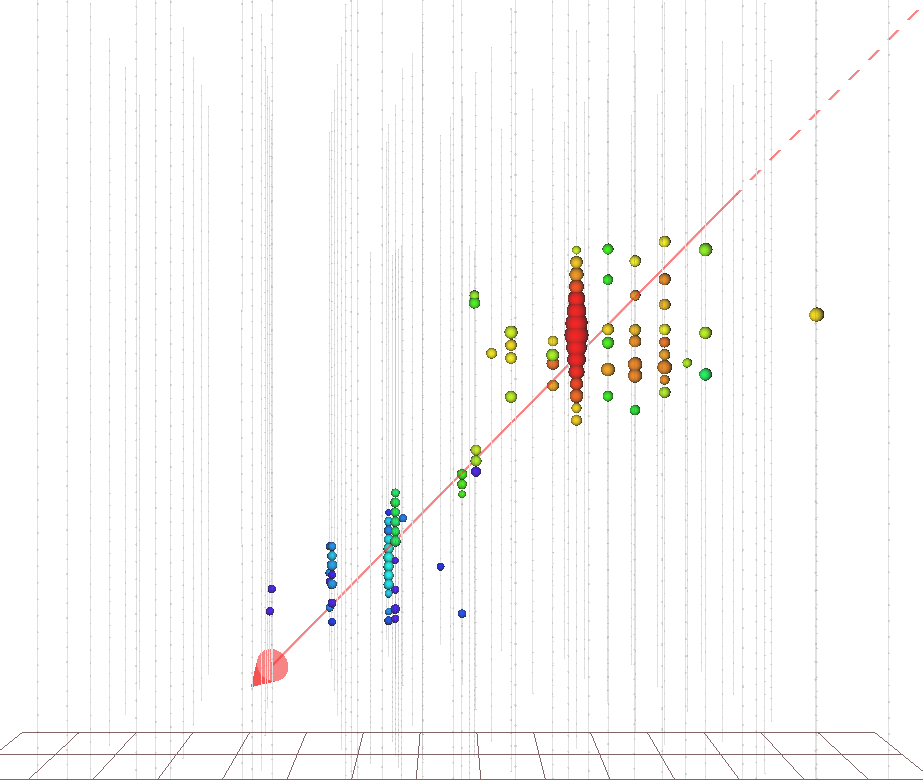}
    \includegraphics[width=0.49\textwidth]{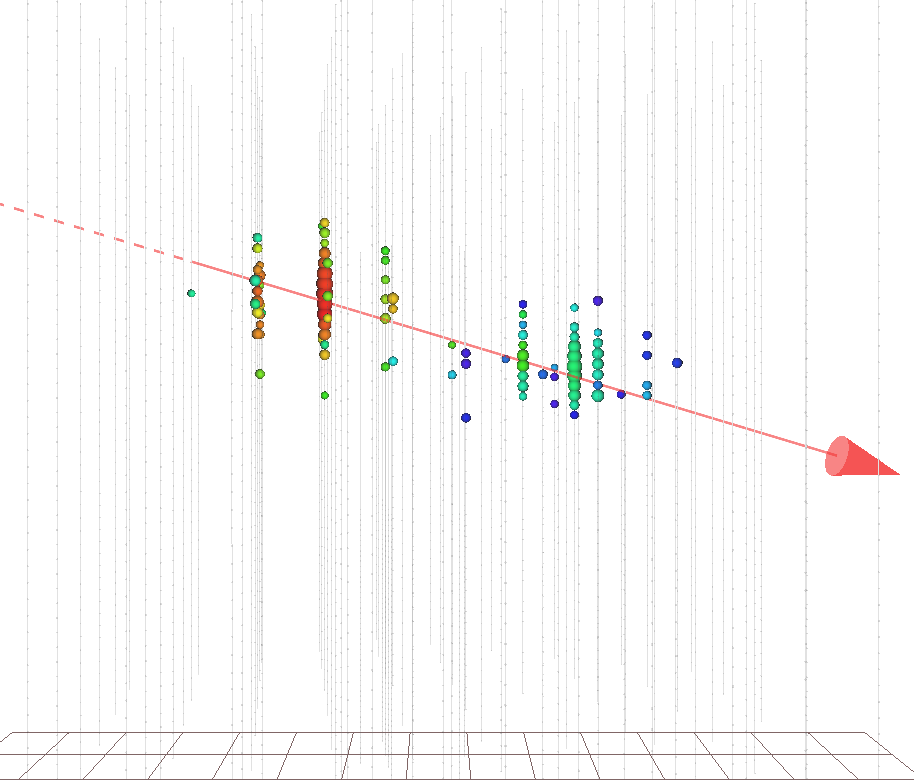}
    \caption{Display of two archival data events in the selection. (Left / Right): Reconstructed energy = 7.3 TeV / 9.8 TeV, reconstructed zenith = 53.6$^\circ$ / 73.5$^\circ$, and signalness = 0.295 / 0.156 according to the sigmoid fit parameterization (see Eq. \ref{eq:Sigmoid fit}).}
    \label{fig:example event views}
\end{figure}

The dominant event morphology in this selection is starting tracks, generated predominantly by charged current muon neutrino interactions in the ice inside the detector's volume. Two examples are shown in Fig. \ref{fig:example event views}. The zenith angle ($\theta$) is in IceCube detector coordinates. $\theta$ = 0 means coming from the southern sky, up towards the Earth, and $\theta$ = $\pi$ means coming from the northern sky, down through the Earth.

The event selection has two stages. The first is a series of computationally cheap cuts that are run in real-time on events at the South Pole to greatly reduce event rates. The second is a more computationally expensive set of cuts known as the Enhanced Starting Track Event Selection (ESTES), a high purity muon neutrino sample. Variations of ESTES with slightly different sets of cuts have been used for a neutrino source search \cite{ESTES_NS_ICRC_2019} and for a measurement of the diffuse astrophysical flux \cite{ESTES_Diffuse_ICRC_2019,ESTES_Diffuse_ICRC_2021,ESTES_Diffuse_ICRC_2023,ESTES_Diffuse_Paper}.

\subsection{At South Pole Selection}\label{sec2.1}

The following cuts are run at the South pole in sequential order to cut down event rates. These cuts make use of a directional reconstruction algorithm that is run at the South Pole, which is computationally cheap. Later on in the event selection, it is superseded by a more computationally intensive and effective directional reconstruction algorithm taken from ESTES (see the reconstructed cosine zenith in Figs. \ref{fig:energy and cosine zenith data/mc}, \ref{fig:SAR and FAR}, \ref{fig:2d hists - event rate and binned signalness} and \ref{fig:sigmoid fits and 2d hist sigmoid signalness}). The two algorithms do not always agree, especially for more poorly reconstructed events, where one algorithm may say it comes from the southern sky and the other may say it comes from the northern sky.

Each event must have:
\begin{enumerate}
    \item At least 450 photoelectrons detected across all DOMs after the reconstructed interaction vertex time
    \item Reconstructed zenith angle $\theta_{Pole}$ < 75$^\circ$
    \begin{enumerate}
    \item Different reconstruction algorithm from that shown in Figs. \ref{fig:energy and cosine zenith data/mc}, \ref{fig:SAR and FAR}, \ref{fig:2d hists - event rate and binned signalness} and \ref{fig:sigmoid fits and 2d hist sigmoid signalness}. Some events are miss-reconstructed at the South Pole (passing this $\theta_{Pole}$ < 75$^\circ$ cut) and are later reconstructed by ESTES to be from the northern sky instead.
    \end{enumerate}
    \item Probability of an event being an incoming muon < 0.001
    \begin{enumerate}
    \item Derived from ESTES's Starting Track Veto, see \textsection\ref{sec2.2}
    \end{enumerate}
    \item Length of the track inside the detector > 400 m
\end{enumerate}

Events that pass all of these cuts will then be sent north and run through ESTES, which takes several minutes.

\subsection{Enhanced Starting Track Event Selection}\label{sec2.2}

The ESTES event selection has been detailed in previous and concurrent ICRC contributions (see Ref. \cite{ESTES_Diffuse_ICRC_2019, ESTES_NS_ICRC_2019, ESTES_Diffuse_ICRC_2021, ESTES_Diffuse_ICRC_2023}). The key features are the Starting Track Veto, a cut which uses an event's reconstructed direction to define a dynamic veto region to heavily reduce atmospheric muon rates (see Fig. 1 in Ref. \cite{ESTES_Diffuse_ICRC_2021} and a Boosted Decision Tree classifier, which further reduces background rates. The remaining events form a high purity muon neutrino sample, dominated by southern sky neutrinos.

A measurement of the diffuse astrophysical neutrino flux using ESTES is detailed in a concurrent ICRC proceeding (see Ref. \cite{ESTES_Diffuse_ICRC_2023}). The Monte Carlo and data shown here are a direct subset of those shown in the ESTES diffuse analysis. This event selection makes use of the same variation of ESTES cuts used for the diffuse analysis. The addition of the cuts done at Pole make this subset have a far higher astrophysical neutrino purity than ESTES.

\subsection{Expected Final Event Rates}\label{sec2.3}

To estimate the rate of astrophysical neutrino events, we use Monte Carlo and the results of the ESTES diffuse single power law fit (see Ref. \cite{ESTES_Diffuse_ICRC_2023}) for the astrophysical neutrinos, $\phi_{\mathrm{Astro}}^{\mathrm{per-flavor}}$ = 1.68 and $\gamma_{\mathrm{Astro}}$ = 2.58 and their systematic uncertainties. To estimate the rate of atmospheric neutrino events, we used the GaisserH4a \cite{GaisserH4a} cosmic ray model and the Sibyll2.3c \cite{Sibyll2.3C} hadronic interaction model. Error bars on the expected rates are produced by sampling the parameters in the 68\% error contour in $\phi_{\mathrm{Astro}}^{\mathrm{per-flavor}}$ vs $\gamma_{\mathrm{Astro}}$ space provided by the ESTES diffuse single power law fit \cite{ESTES_Diffuse_ICRC_2023,ESTES_Diffuse_Paper}.

There are two variables of relevance: the reconstructed neutrino energy and the reconstructed cosine zenith, both of which are the same as used in the ESTES diffuse analysis. The reconstructed energy comes from the ESTES Random Forest algorithm (based off Ref. \cite{RandomForest}). The reconstructed cosine zenith comes from the ESTES diffuse reconstruction algorithm using Millipede (see Ref. \cite{ESTES_Diffuse_ICRC_2023}). 

\begin{figure}[h!]
    \centering
    \includegraphics[width=0.49\textwidth]{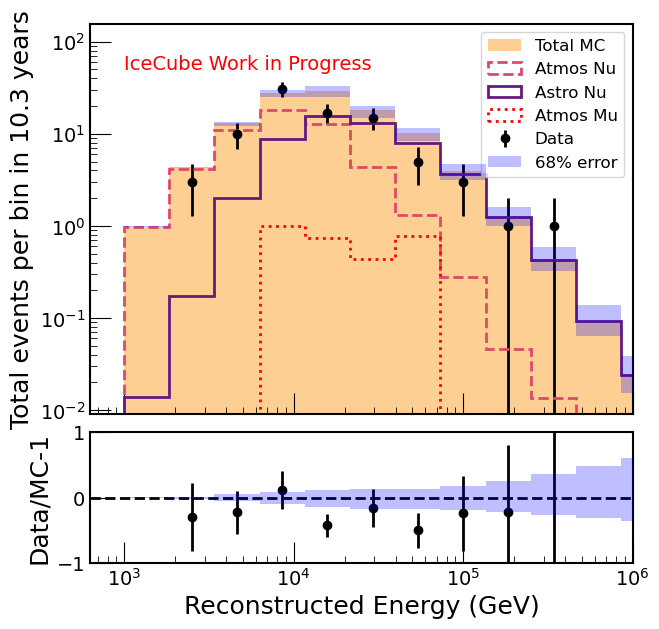}
    \includegraphics[width=0.49\textwidth]{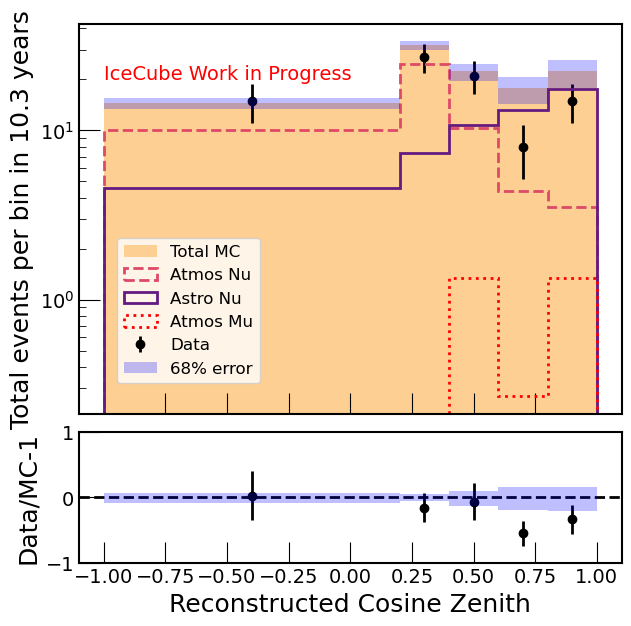}
    \caption{Data vs Monte Carlo comparisons for the energy and cosine zenith variables, which are used to define the signalness on a per event basis. The MC over predicts the data substantially in 2 energy bins and 1 cosine zenith bin, and work is in progress to resolve this.}
    \label{fig:energy and cosine zenith data/mc}
\end{figure}

The data/MC plots for these variables are shown in Fig. \ref{fig:energy and cosine zenith data/mc}. The reconstructed energy shown here will also be sent out in the real-time alert. The reconstructed cosine zenith shown here will not be sent out in the real-time alert; a more robust method will be used instead (see \textsection\ref{sec4}).

There is currently an observed normalization issue where the MC over predicts the number of events compared to the data, which is most prominent in two energy bins and one cosine zenith bin in Fig. \ref{fig:energy and cosine zenith data/mc}. The ESTES diffuse event selection observes good data/MC agreement in both of these parameters across the parameter space (see Fig. 3 in Ref. \cite{ESTES_Diffuse_ICRC_2023}), so this tension is derived solely from the cuts unique to the real-time process at Pole. Resolving this discrepancy is underway.

\begin{figure}[h!]
    \centering
    \includegraphics[width=0.49\textwidth]{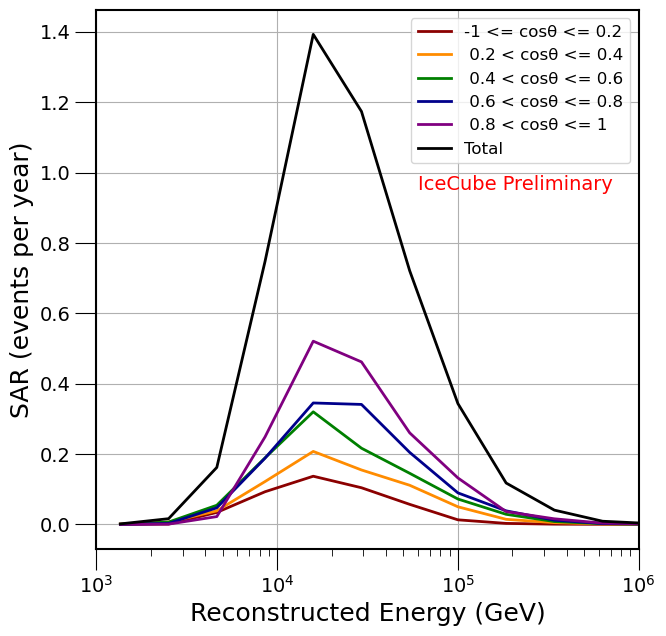}
    \includegraphics[width=0.49\textwidth]{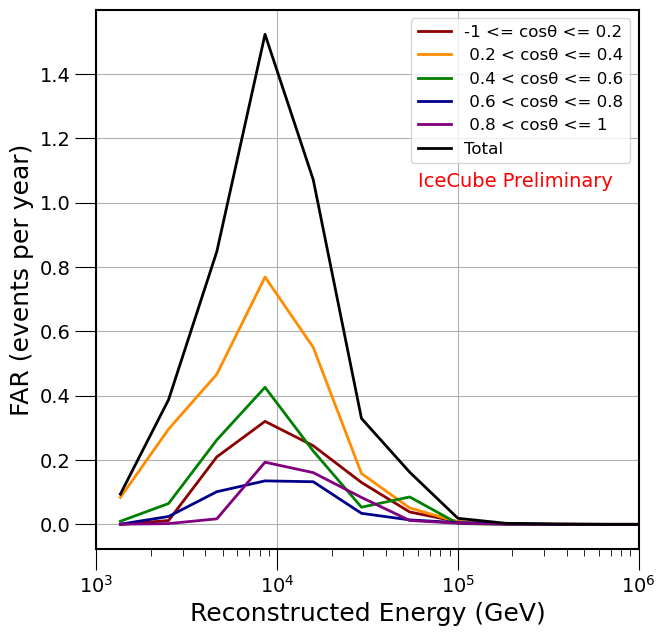}
    \caption{Left: Signal Alarm Rate vs energy, for each of the cosine zenith bins (total = 5.1 events per year). Right: False Alarm Rate vs energy for different cosine zenith bins (total = 5.2 events per year). The energy bin widths are the same as in Fig. \ref{fig:energy and cosine zenith data/mc}.}
    \label{fig:SAR and FAR}
\end{figure}

The reconstructed energy and cosine zenith are used to build our event rate expectations, as seen in Fig. \ref{fig:SAR and FAR}. The Signal Alarm Rate (SAR) defines how many genuine astrophysical events will be sent out as alerts per year, and the False Alarm Rate (FAR) defines how many background events will be sent out as alerts per year.

In total, we expect 10.3 events per year to be sent out as alerts, 50\% of which will be of astrophysical origin. This is the same purity standard as IceCube Gold real-time alerts, so this selection can be added directly into the current real-time stream. In looking at 10.3 years of archival data, there has been no observed overlap between this selection and any of IceCube's current real-time event selections (GFU, HESE and EHE in \cite{IceCubeRealtime}). 

\section{Signalness Calculation}\label{sec3}

\begin{figure}[h!]
    \centering
    \includegraphics[width=0.49\textwidth]{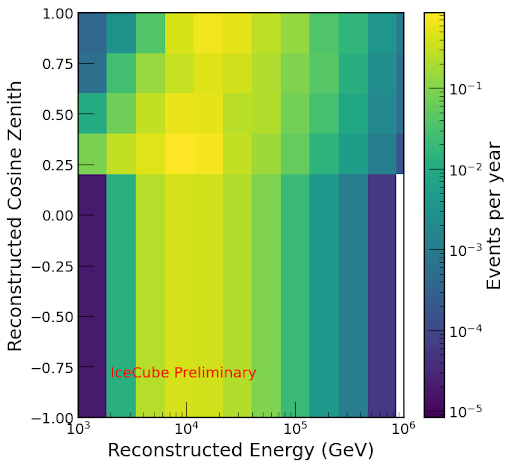}
    \includegraphics[width=0.49\textwidth]{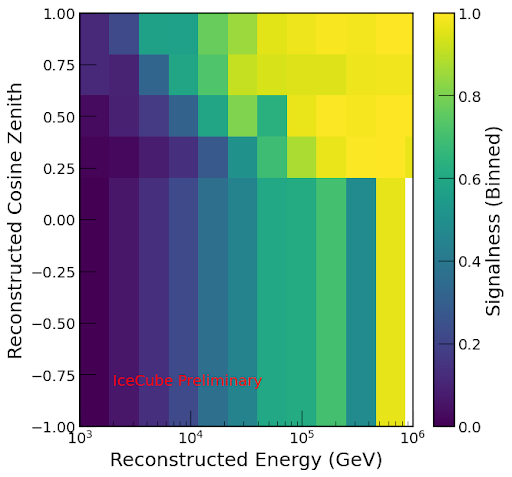}
    \caption{2d histograms of the event rates (same as both panels of Fig. \ref{fig:SAR and FAR} added together) for this event selection (left) and the signalness of events in a given bin (right) binned in the reconstructed energy and reconstructed cosine zenith.}
    \label{fig:2d hists - event rate and binned signalness}
\end{figure}

Because these events are sent out in real-time to other physics experiments, it is useful to quantify how interesting an event is with \textit{signalness}, the probability that the event is astrophysical. We calculate the signalness on an event-by-event basis, first by taking our MC and splitting it into different energy and cosine zenith bins according to Fig. \ref{fig:energy and cosine zenith data/mc}, which can be seen in Fig. \ref{fig:2d hists - event rate and binned signalness}. Formally, 

\begin{equation}
    Signalness(E,\theta) = \frac{N_{astro}(E,\theta)}{N_{astro}(E,\theta) + N_{bkg}(E,\theta)},
    \label{eq:Binned signalness}
\end{equation}

where $N_{astro}$ is the "Astro Nu" and $N_{bkg}$ is the "Atmos Nu" and "Atmos Mu" as seen in Fig. \ref{fig:energy and cosine zenith data/mc} across the reconstructed energy and reconstructed cosine zenith space.

Then, we fit each reconstructed cosine zenith bin with a 1d sigmoid function in reconstructed energy space to interpolate $S(E, \theta)$ on an event by event basis. The sigmoid fit function for each reconstructed cosine zenith bin has 3 free parameters, and is given by

\begin{equation}
    Signalness = \frac{1-c}{1+e^{-k_{o}(log_{10}(E)+x_{o})}} + c.
    \label{eq:Sigmoid fit}
\end{equation}

\begin{figure}[h!]
    \centering
    \includegraphics[width=0.49\textwidth]{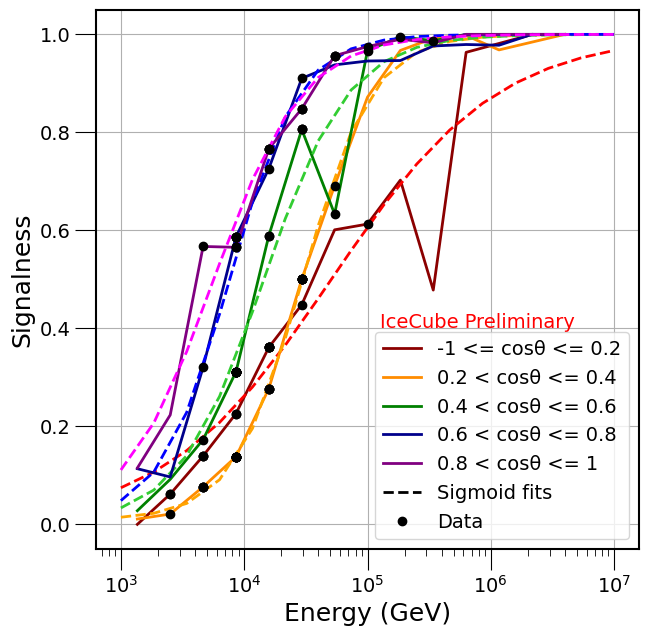}
    \includegraphics[width=0.49\textwidth]{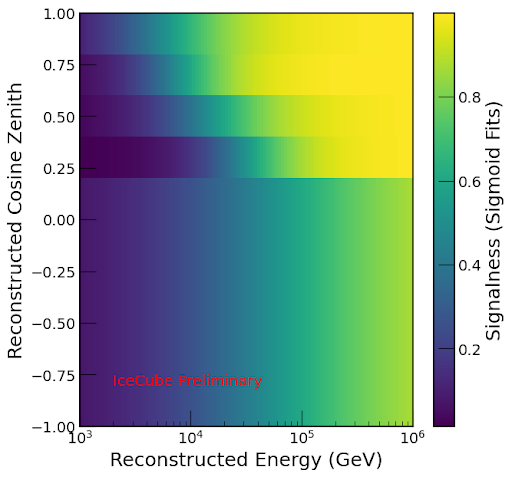}
    \caption{Left: 1d sigmoid fits to each of the cosine zenith bins shown in Fig. \ref{fig:2d hists - event rate and binned signalness} across the energy range. Black data points represent cosine zenith/energy bins that contained archival data. Right: 2d histogram of the signalness of events according to the sigmoid fit parameterization.}
    \label{fig:sigmoid fits and 2d hist sigmoid signalness}
\end{figure}

The sigmoid fits and their resultant signalness values can be seen in Fig. \ref{fig:sigmoid fits and 2d hist sigmoid signalness}.

\begin{figure}[h!]
    \centering
    \includegraphics[width=0.49\textwidth]{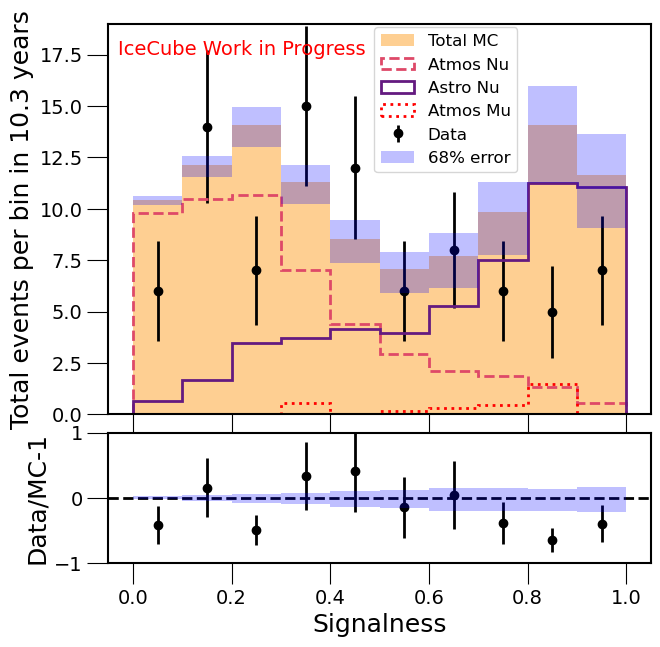}
    \caption{Data vs Monte Carlo comparison for the signalness evaluated by Eq. \ref{eq:Sigmoid fit}. The MC over predicts the data substantially in 3 signalness bins, a direct propagation of the data/MC mismatch seen in Fig. \ref{fig:energy and cosine zenith data/mc}. Work is in progress to resolve this.}
    \label{fig:signalness data/mc}
\end{figure}

The final signalness distribution evaluated by Eq. \ref{eq:Sigmoid fit} is shown in Fig. \ref{fig:signalness data/mc}, and can be directly compared with the distribution of current alerts seen in Fig. 5 of Ref. \cite{IceCubeRealtime}. This is calculated based off of the reconstructed energy and cosine zenith on a per event basis according to the sigmoid fit parameterization. This variable will be sent out in the real-time alert.

Here, we see that the events peak around a signalness of 0.2 and 0.8 and trough in the middle around 0.55.

\section{Real-time Alert Stream}\label{sec4}

All events that pass the event selection will be sent out as alerts in real-time, after running a final directional reconstruction algorithm, which performs a likelihood scan assuming the event originated from each pixel of the sky, iterating to more granular pixels as it approaches the best fit position (see Fig. 2 in Ref. \cite{Skymap_Scanner_ICRC_2023}). The information sent to the multi-messenger community in the alert will be:

\begin{enumerate}
    \item Timing information for the event
    \item Reconstructed best fit direction in right ascension and declination evaluated by a likelihood scan of the entire sky
    \begin{enumerate}
    \item Expect a median space angle resolution of 1.5$^\circ$ for alert events
    \end{enumerate}
    \item Uncertainties on the right ascension and declination taken from the bounding rectangle on the 90\% containment angular error contour, including a link to the likelihood skymap
    \begin{enumerate}
    \item Expect median error in declination of $\pm$1.3$^\circ$ and a median error in right ascension of $\pm$3$^\circ$
    \end{enumerate}
    \item Reconstructed energy of the event, taken from the ESTES Random Forest algorithm
    \begin{enumerate}
    \item Energy resolution of 25\% - 30\% between 1 TeV and 10 PeV (see \cite{ESTES_Diffuse_ICRC_2023})
    \item Based off Ref. \cite{RandomForest}
    \end{enumerate}
    \item Signalness of the event, calculated as described above (sigmoid fits)
    \begin{enumerate}
    \item Expect an average event signalness of 50\% (IceCube Gold)
    \end{enumerate}
    %\begin{enumerate}
    %\end{enumerate}
\end{enumerate}    

There is currently ongoing work to refine the procedure for generating angular error contours for events sent out as alerts for this selection. This is due to observed under-coverage in our MC studies (X\% containment contours cover the true event direction < X\% of the time). This work may result in an additional cut on the event selection to remove events that are not well localized, or increased uncertainties on the right ascension and declination that get sent out as alerts.

\section{Conclusion}\label{sec5}

Current estimates from the MC tell us that we expect 10.3 events per year from this alert stream. Given the fact that there has been no observed overlap at the alert level between this alert stream and the existing IceCube alert streams, these events would be unique additions to IceCube Gold, and would roughly double the number of IceCube Gold alerts sent out per year. These events would also open up more of the parameter space to the community, largely coming from the southern sky and at lower energies (mostly 5-100 TeV). This work is still undergoing development before being added to the IceCube real-time stream.

% Bibtex references:
\bibliographystyle{ICRC}
\bibliography{references}

% Alternatively, you can include references by hand:
%\begin{thebibliography}{99}
%\bibitem{...}
%
%\end{thebibliography}

\clearpage

%The following list of authors, affiliations and funding agencies will be updated at the day of submission. The following template is a placeholder generated via https://authorlist.icecube.wisc.edu/icecube on March 25, 2023 and will be updated.
\input{authorlist_IceCube.tex}

\end{document}

%% file: authorlist_IceCube.tex
\section*{Full Author List: IceCube Collaboration}

\scriptsize
\noindent
R. Abbasi$^{17}$,
M. Ackermann$^{63}$,
J. Adams$^{18}$,
S. K. Agarwalla$^{40,\: 64}$,
J. A. Aguilar$^{12}$,
M. Ahlers$^{22}$,
J.M. Alameddine$^{23}$,
N. M. Amin$^{44}$,
K. Andeen$^{42}$,
G. Anton$^{26}$
% ...
\\
\\
% ...
$^{12}$ Universit{\'e} Libre de Bruxelles, Science Faculty CP230, B-1050 Brussels, Belgium \\
% ...
$^{17}$ Department of Physics, Loyola University Chicago, Chicago, IL 60660, USA \\
$^{18}$ Dept. of Physics and Astronomy, University of Canterbury, Private Bag 4800, Christchurch, New Zealand \\
% ...
$^{22}$ Niels Bohr Institute, University of Copenhagen, DK-2100 Copenhagen, Denmark \\
$^{23}$ Dept. of Physics, TU Dortmund University, D-44221 Dortmund, Germany \\
% ...
$^{26}$ Erlangen Centre for Astroparticle Physics, Friedrich-Alexander-Universit{\"a}t Erlangen-N{\"u}rnberg, D-91058 Erlangen, Germany \\
% ...
$^{40}$ Dept. of Physics and Wisconsin IceCube Particle Astrophysics Center, University of Wisconsin{\textendash}Madison, Madison, WI 53706, USA \\
% ...
$^{42}$ Department of Physics, Marquette University, Milwaukee, WI, 53201, USA \\
% ...
$^{44}$ Bartol Research Institute and Dept. of Physics and Astronomy, University of Delaware, Newark, DE 19716, USA \\
% ...
$^{63}$ Deutsches Elektronen-Synchrotron DESY, Platanenallee 6, 15738 Zeuthen, Germany  \\
$^{64}$ Institute of Physics, Sachivalaya Marg, Sainik School Post, Bhubaneswar 751005, India \\
% ...

\subsection*{Acknowledgements}

\noindent
The authors gratefully acknowledge the support from the following agencies and institutions:
USA {\textendash} U.S. National Science Foundation-Office of Polar Programs,
U.S. National Science Foundation-Physics Division,
U.S. National Science Foundation-EPSCoR,
Wisconsin Alumni Research Foundation,
Center for High Throughput Computing (CHTC) at the University of Wisconsin{\textendash}Madison,
Open Science Grid (OSG),
Advanced Cyberinfrastructure Coordination Ecosystem: Services {\&} Support (ACCESS),
Frontera computing project at the Texas Advanced Computing Center,
U.S. Department of Energy-National Energy Research Scientific Computing Center,
Particle astrophysics research computing center at the University of Maryland,
Institute for Cyber-Enabled Research at Michigan State University,
and Astroparticle physics computational facility at Marquette University;
Belgium {\textendash} Funds for Scientific Research (FRS-FNRS and FWO),
FWO Odysseus and Big Science programmes,
and Belgian Federal Science Policy Office (Belspo);
Germany {\textendash} Bundesministerium f{\"u}r Bildung und Forschung (BMBF),
Deutsche Forschungsgemeinschaft (DFG),
Helmholtz Alliance for Astroparticle Physics (HAP),
Initiative and Networking Fund of the Helmholtz Association,
Deutsches Elektronen Synchrotron (DESY),
and High Performance Computing cluster of the RWTH Aachen;
Sweden {\textendash} Swedish Research Council,
Swedish Polar Research Secretariat,
Swedish National Infrastructure for Computing (SNIC),
and Knut and Alice Wallenberg Foundation;
European Union {\textendash} EGI Advanced Computing for research;
Australia {\textendash} Australian Research Council;
Canada {\textendash} Natural Sciences and Engineering Research Council of Canada,
Calcul Qu{\'e}bec, Compute Ontario, Canada Foundation for Innovation, WestGrid, and Compute Canada;
Denmark {\textendash} Villum Fonden, Carlsberg Foundation, and European Commission;
New Zealand {\textendash} Marsden Fund;
Japan {\textendash} Japan Society for Promotion of Science (JSPS)
and Institute for Global Prominent Research (IGPR) of Chiba University;
Korea {\textendash} National Research Foundation of Korea (NRF);
Switzerland {\textendash} Swiss National Science Foundation (SNSF);
United Kingdom {\textendash} Department of Physics, University of Oxford.